\documentclass[submission,copyright,creativecommons]{eptcs}
\providecommand{\event}{Refine 2015} 

\usepackage{color}
\usepackage{url}
\usepackage[latin1]{inputenc}
\usepackage{amsmath}
\usepackage{amscd}
\usepackage{latexsym}
\usepackage{stmaryrd}
\usepackage{amssymb}
\usepackage{amscd}
\usepackage{fancyvrb}
\usepackage{boxedminipage}

\ifpdf
\usepackage[pdftex]{graphicx}
\else
\usepackage{graphicx}
\fi
 
 \usepackage[all]{xy}
\newtheorem{definition}{Definition}[section]
\newtheorem{example}{Example}[section]
\newtheorem{theorem}{Theorem}[section]

\newtheorem{lemma}{Lemma}[section]

\newtheorem{corollary}{Corollary}[section]

  \def\HHL{\mathcal{HHL}}
 
\def\sat#1{\models_{#1}}

\newcommand{\Nom}{\mathrm{Nom}}
\newcommand{\Prop}{\mathrm{Prop}}

\newcommand{\Mod}{\mathrm{Mod}}

\newcommand{\REL}{\REL}

\def\PL{\mathit{PL}}
\def\REL{\mathit{REL}}

\def\H2PL{\mathcal{H}^2\PL}

\newcommand{\s}{s}

\def\just#1#2{\\
         &#1& \rule{2em}{0pt} \footnotesize{ \{ \mbox{\rule[-.7em]{0pt}{1.8em} #2} \}} \\ && }

\newcommand{\refinesl}[2]{{#1}\rightsquigarrow_l {#2}}
\newcommand{\refinesh}[2]{{#1}\rightsquigarrow_h {#2}}

\def\rcb#1#2#3#4{\def\nothing{}\def\range{#3}\mathopen{\langle}#1 \ #2 \ \ifx\range\nothing::\else: \ #3 :\fi \ #4\mathclose{\rangle}}

\def\just#1#2{\\
         &#1& \rule{2em}{0pt} \footnotesize{ \{ \mbox{\rule[-.7em]{0pt}{1.8em} #2} \}} \\ && }

\DeclareMathAlphabet{\mathbb}{U}{msb}{m}{n}
\DeclareSymbolFont{ams}{U}{msa}{m}{n}
\DeclareSymbolFontAlphabet{\mathams}{ams}
\DeclareMathSymbol{\filter}{\mathams}{ams}{22}

\allowdisplaybreaks
\newcommand{\ST}{\mathrm{ST}}
\newcommand{\HR}{\rightharpoonup}

\newcommand{\qed}{\hspace*{\fill}$\blacksquare$}

\title{A logic for n-dimensional  hierarchical refinement}
\author{
Alexandre Madeira 
\institute{HASLab - INESC TEC \& Univ. Minho, \\ Braga, Portugal} 
\email{amadeira@inesctec.pt} 
\and
Manuel A. Martins 
\institute{CIDMA - Dep. Mathematics, \\ Univ. Aveiro,  Portugal} 
\email{martins@ua.pt} 
\and
Lu\' is S.~Barbosa
\institute{HASLab - INESC TEC \& Univ. Minho \\ Braga, Portugal}
\email{lsb@di.uminho.pt}
}
\def\titlerunning{A logic for n-dimensional hierarchical refinement}
\def\authorrunning{A. Madeira, M.A. Martins \& L.S. Barbosa}
\begin{document}
 \maketitle

 \begin{abstract}
Hierarchical transition systems provide a popular mathematical structure to represent state-based software applications in which different layers of abstraction are represented by inter-related state machines. The decomposition of high level states into inner sub-states, and of their transitions into inner sub-transitions is common refinement procedure adopted in  a number of specification formalisms. 

This paper introduces a hybrid modal logic for k-layered transition systems, its first-order standard translation, a notion of bisimulation, and a modal invariance result.  Layered and hierarchical notions of refinement are also discussed in this setting.
\end{abstract}

\section{Motivation and aims} 
\label{sec:motivation}

\begin{figure}[ht!]
\begin{center}
   \includegraphics[width=0.6\textwidth,natwidth=610,natheight=642]{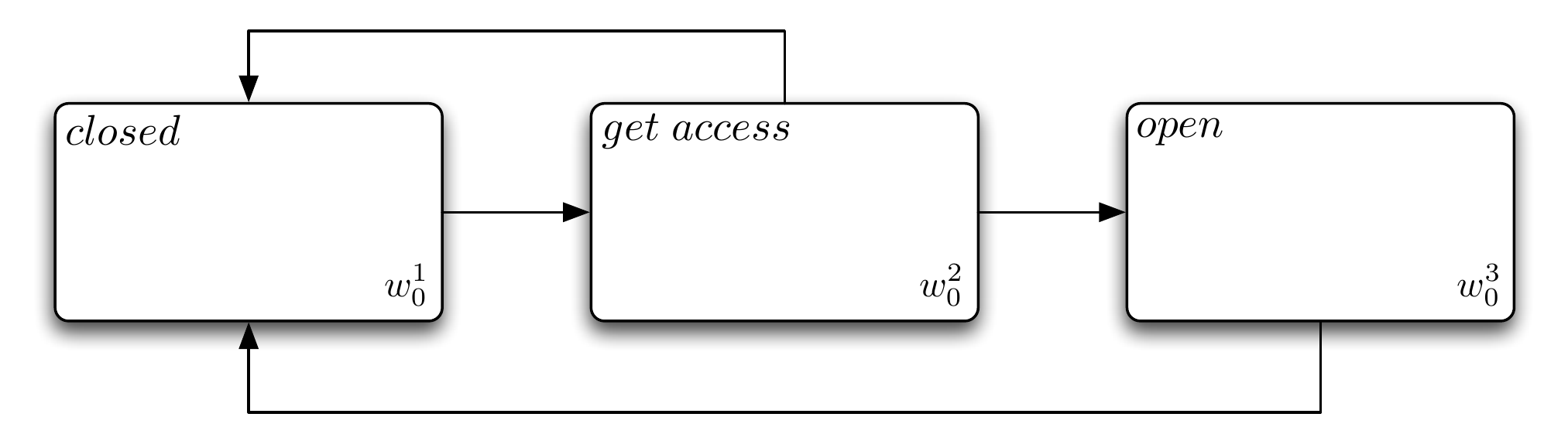}
\end{center}
\caption{An abstract strongbox behavioural model.}
\label{fig:intro1}
\end{figure}

Figure \ref{fig:intro1} depicts a high level behavioural model of a strongbox controller in the form of a transition system with three states. The strongbox can be 
open, closed, or going through an authentication process. The model can be formalised in some sort of modal logic, so that state transitions can be expressed, possibly combined with hybrid features to refers to specific, individual states. 
Recall that the  qualifier \emph{hybrid} \cite{manifesto} applies to extensions of modal languages with symbols, called \emph{nominals}, which explicitly
refer to individual states in the underlying Kripke frame. A satisfaction operator $@_i \varphi$ is included standing for $\varphi$ holding in the state named by nominal $i$.
For example, in propositional hybrid logic \cite{livro_brauner} and assuming a set of nominals $\Nom=\{closed, get\; access, open\}$, we can express the dynamics depicted in the diagram of Figure \ref{fig:intro1}, \emph{e.g.},
\begin{itemize}
\item  that the state $get\;access$ is accessible from the state $closed$, with $@_{closed} \Diamond get\;access$, or
\item that the state $open$ is not directly accessible from $closed$, with $\Diamond open \rightarrow \neg closed$.
\end{itemize}

\begin{figure}[ht!]
\begin{center}
 \includegraphics[width=0.8\textwidth,natwidth=610,natheight=642]{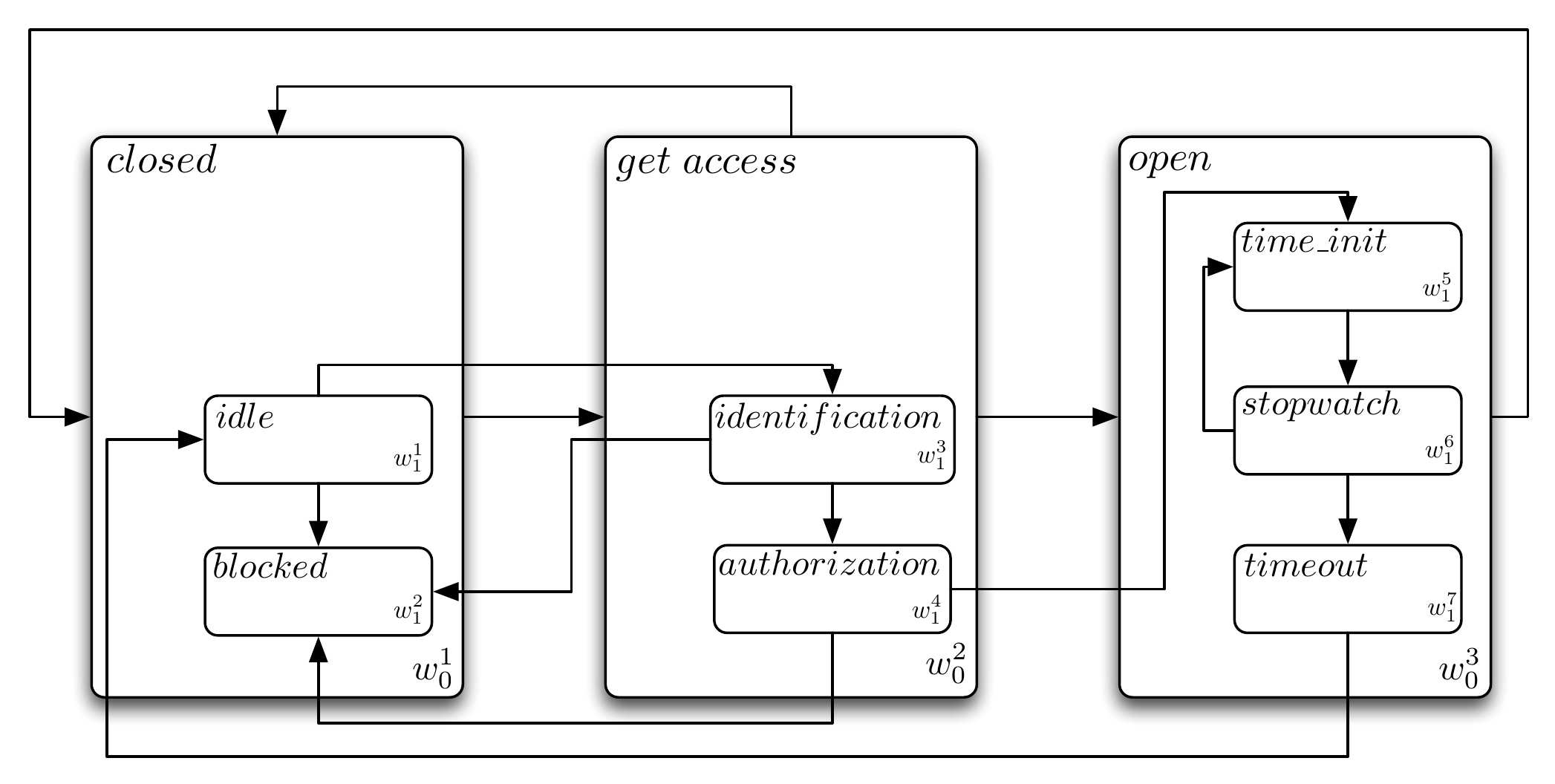}
\end{center}
\caption{A 2-layered refined strongbox model.}
\label{fig:intro2}
\end{figure}

\noindent 
This  high level vision of the strongbox controller can be refined by decomposing not only its internal states,  but also its transitions. Thus, each `high-level' state gives rise to a new, local transition system, and each  `high-level'-transition is decomposed into a number of `intrusive' transitions from  sub-states of the  `down level'-transition system corresponding to the refinement of the original source state, to sub-states of the corresponding refinements of original target states.  For instance, the (upper) $close$ state can be refined into  a  (inner) transition system with two (sub) states, one, $idle$,
 representing the system waiting for the order  to proceed for the $get\;access$ state and, another one, $blocked$, capturing a system which is unable to proceed with the opening process (e.g. when  authorised access for a given user was definitively denied). In this scenario, the upper level transition
from  $closed$ to $get\;access$  can be realised by, at least, one intrusive transition between the $closed$ sub-state $idle$ and the $get\;access$ sub-state $identification$ where the user identification to proceed is supposed to be checked. Figure~\ref{fig:intro2} illustrates the result of this refinement step. 

\begin{figure}[ht!]
\begin{center}
   \includegraphics[width=0.3\textwidth,natwidth=610,natheight=642]{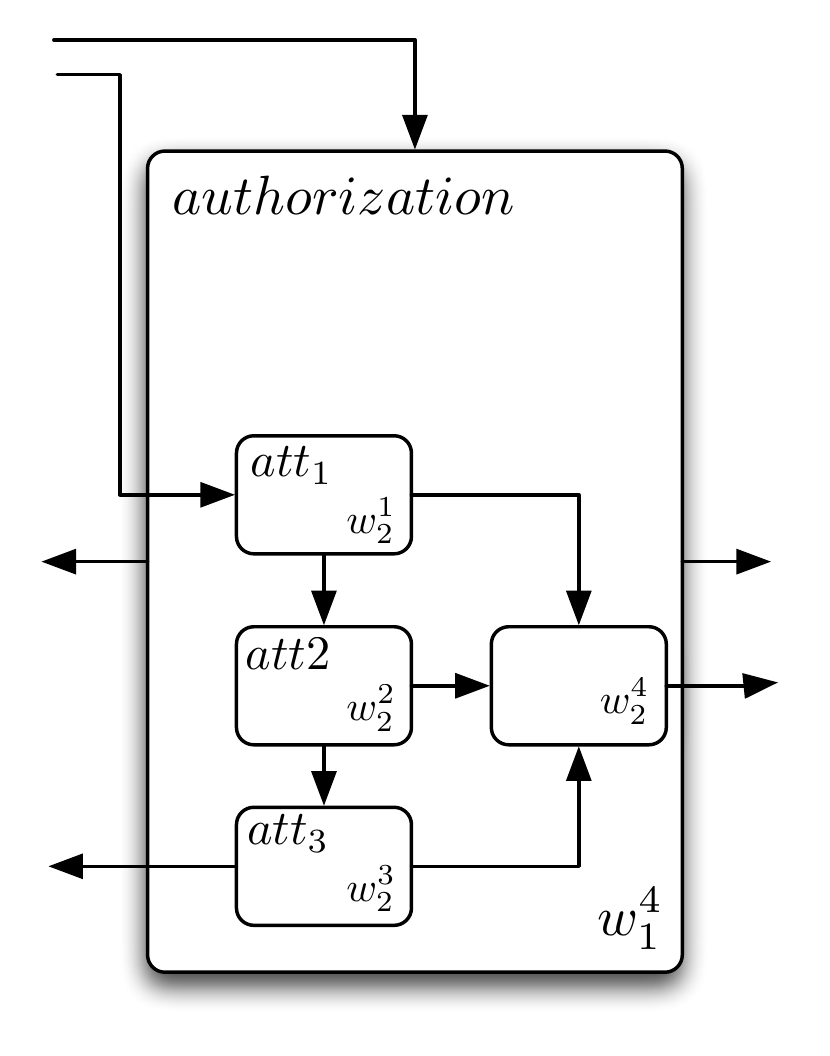}
\end{center}
\caption{Fragment of the 3-layered refined strongbox model.}
\label{fig:intro3}
\end{figure}

\noindent 
Still the specifier may go even further. For example, he may like to refine the $get\; access$ sub-state $authorisation$ into the more fine-grained transition structure depicted in Figure~\ref{fig:intro3}. This third-level view includes  a sub-state corresponding to each one of the possible three attempts of password validation, as well as an auxiliary state to represent the authentication success.  

Such an  hierarchical way to design a system is quite natural and somehow  inherent to well known design formalisms such as David  Harel's statecharts \cite{DBLP:journals/scp/Harel87} and the subsequent UML hierarchical state-machines \cite{citeulike:709834}, and action refinement \cite{DBLP:journals/acta/GlabbeekG01}, among others. 
 
This paper introduces a hierarchical hybrid logic in order to  express, and reason about, requirements which typically involve transitions between designated states in different local transition systems, such as, for example,  the ones designated by $identification$ and $blocked$ in Figure~\ref{fig:intro2}. This extends our previous work \cite{aiml14} on hierarchical logic in order to capture truly intrusive transitions which are required to express complex software designs as described \emph{e.g.} with statecharts. Suitable notions of bisimulation, and corresponding invariance results, as well as layered and hierarchical refinement are introduced and illustrated.
%
%

The paper is organised as follows: Section \ref{sc:logic} introduces the logic, whose basic modal theory, inclu\-ding a standard translation to first-order logic and a modal invariance result, is discussed in section \ref{sc:meh}. Layered and hierarchical refinements are considered in section  \ref{sc:refine}. Finally, section   \ref{sc:tim} concludes the paper and points out some current developments.

\section{The logic}\label{sc:logic}
This section introduces a multi-layer hybrid logic to reason upon hierarchical transition systems. The adoption of the term `hybrid' is addressed to the terminology of the modal logic community (eg. \cite{livro_brauner,manifesto}). The `hybrid' nature of the formalism is regarded to the combination of aspects of first-order and modal logic. Note that this should not be confused with the different, usual, meaning of the the same term to mention systems with mixed continuous and discrete behaviours. 

We start fixing a 
 few notational conventions.   Given a family $A=(A_i)_{i\in \{0,\cdots, n\}}$, we denote by $A[k]$ the sub-family $A[k]=(A_i)_{i\in \{0,\cdots, k\}}$.
 Given a predicate $P\subseteq S_1 \times \cdots \times S_n$ we denote by $P|_k$ the restriction of $P$ to its first $k$ components, i.e. the predicate $P|_k \subseteq  S_1 \times \cdots \times S_k$ such that 
 \[P|_k=\big\{(s_1,\cdots, s_k)|P(s_1,\cdots, s_k, s_{k+1},\cdots,s_n)\mbox{ for some } s_r\in S_r, r\in \{k+1,\cdots,n\}\big\}\] Given a relation  $R \subseteq (S_1 \times \dots \times S_n)^2$, we denote by $R|_k$ the relation $R|_k \subseteq  (S_1 \times \cdots \times S_k)^2$ such that  
 \begin{eqnarray*}
 R|_k=  
 \big\{(s_1,\cdots, s_k, s'_1, \cdots, s'_k)|R(s_1,\cdots,s_k,s_{k+1},\cdots,s_n,s'_1,\cdots s'_k,s'_{k+1},\cdots s'_n)\mbox{ for } \\s_r,s'_r\in S_r, r\in \{k+1,\cdots,n\}\big\}
 \end{eqnarray*}

\noindent
The logic can now be introduced as follows.

  \subsection*{Signatures}
  Signatures are $n$-families of disjoint, possible empty, sets of symbols $$\Delta^n=\big(\Prop_k,\Nom_k\big)_{k\in\{0,\cdots,n\}}$$
  
\begin{example}
  To express the strongbox model introduced above as a running example,  we have to define a signature $\Delta^2$ for the three layers presented. Note that, for sake of simplicity, the level-subscripts of nominals are omitted in the diagrams above.  $0$-level symbols consist of  the set of nominals $\Nom_0=\{closed_0,get\_access_0, open_0\}$, and a set of propositions $\Prop_0$ including, for instance, a proposition $safe\_state_0$ to be assigned to the $0$-states where the machine is not opened. For the $1$-level signature we consider a set of nominals 
  \[\Nom_1=\{idle_1, blocked_1, identification_1, authorization_1, time\_init_1, stopwatch_1, time\_out_1\}\, \]
  and a set of propositions $\Prop_1$ which may include, for example,  a proposition $timed\_state_1$ to be assigned to timed dependent $1$-states (e.g. the inner states of the one named by $open_0$ and the 1-state named by $authorization_1$). The fragment presented in Figure~\ref{fig:intro3} entails the inclusion of nominals ${att_1}_2$, ${att_2}_2$ and ${att_3}_2$ in $\Nom_2$.
\end{example}

\subsection*{Formulas} The set of formulas $Fm(\Delta^n)$ is the $n$-family recursively defined, for each $k$, by

\[\varphi_0\ni i_0\;|\;p_0\;|\; \neg \varphi_0 \;|\; \varphi_0 \wedge \varphi_0\;|\;  @_{i_0} \varphi_0\;|\; \diamond_0 \varphi_0
\]

\[\varphi^b_0\ni i_0\;|\;p_0\;|\;  @_{i_0} \varphi_0\;|\; \diamond_0 \varphi_0
\]

and 
  \[\varphi_k\ni \varphi^b_{k-1}\;|\; i_k\;|\;p_k\;|\; \neg \varphi_k \;|\; \varphi_k \wedge \varphi_k\;|\; @_{i_k} \varphi_k\;|\; \diamond_k \varphi_k
\]
where for any $k\in \{1,\dots, n\}$, the basic formulas are defined by 
\[\varphi^b_{k-1}  \ni i_{k-1}\;|\;p_{k-1}\;|\;  \varphi^b_{k-2} \;|\; @_k \varphi_{k-1}\;|\; \diamond_{k-1} \varphi_{k-1}\] for $k\in \{2,\cdots, n\}$, 
   $p_k\in \Prop_k$ and $i_k\in \Nom_k$.
  
  For each $k\in\{0,\dots, n\}$, the strict $k$-layered formulas $SFm(k,\Delta^n)$ are defined  as the fragments of $Fm(\Delta^n)$ given by the grammar
  \[\varphi_k\ni   p_k \;|\;i_k \;|\; \neg \varphi_k \;|\; \varphi_k \wedge \varphi_k \;|\;@_{i_k} \varphi_k \;|\; \diamond_k \varphi_k \]
 The positive fragments of $Fm(\Delta^n)$ and $SFm(k,\Delta^n)$, i.e.,  the sets of sentences built by the corresponding grammars but excluding  negations, are denoted by $Fm^+(\Delta^n)$ and $SFm^+(k,\Delta^n)$, respectively.

\begin{example}
  This language is able  to express properties of very different natures, some of them easily identifiable in our running example.   For instance, we may express inner-outer relations between named states (e.g. $@_{idle_1}closed_0$ or $@_{{att_1}_2}open_0$) as well as a variety of transitions. Those include, for example, the layered transition  $@_{get\_access_0} \diamond_0 open_0$, the $0$-internal transition $@_{identification_1} \diamond_1 authorisation_1$ or the $0$-intrusive transitions $@_{idle_1} \diamond_1 authorisation_1$ 
  and  $get\_ access_0 \rightarrow \diamond_1 open_0$.
  Example~\ref{ex:counter} provides further examples of the logic expressiveness.

\end{example}

\subsection*{Models}

\begin{definition}[$n$-layered models]
  A $n$-layered model $M\in \Mod^n(\Delta^n)$ is a tuple
      \[M=(W^n,D^n,R^n,V^n)\]
      recursively defined as follows: 
      
      \begin{itemize}
          \item $W^n=(W_k)_{k\in \{0,\cdots,n\}}$ is a family of disjoint sets
          \item $D^n\subseteq W_0 \times \dots \times W_{n}$ is a predicate 
          such that,  denoting by $D_k$ the $k$-restriction $D^n|_k$, for  each $k\in \{0,\cdots, n\}$, verifies
        $$W_k=\{v_k|D_k(w_0,\cdots,w_{k-1},v_k), \mbox{ for some } w_0,\cdots,w_{k-1} \mbox{ such that }  D_{k-1}(w_0,\cdots,w_{k-1})\}$$

          \item $R^n=\big(R_{k}\subseteq D_k \times D_k)_{k\in \{0,\cdots, n\}}$ is a $n$-family of binary relations;
            
          

             \item
             $V^n=(V^\Prop_k,V^\Nom_k)_{k\in \{0,\cdots, n\}}$ is a family of pairs of functions 
             \begin{itemize}
             \item $V^\Prop_0: \Prop_0 \rightarrow \mathcal{P}(W_0)$ and $V^\Prop_k: \Prop_k\times D_{k-1} \rightarrow \mathcal{P}(W_k)$ for any $k>0$; and 
              \item $V^\Nom_k: \Nom_k \rightarrow W_k $.
             \end{itemize}
             
        \end{itemize} 
\end{definition}

For each $k\in \{0,\cdots n\}$, model $M_k=(W^n[k],D^n|_k,R^n[k],V^n[k])$, is said the $k$-restriction of  $M=\Mod^n(\Delta^n)$.

\medskip\noindent A specific, particularly well-behaved class of layered models, very important in refinement situation, is defined as follows:
\begin{definition}[Hierarchical Model]

  A $n$-layered model $M=(W^n,D^n,R^n,V^n) \in \Mod^n(\Delta^n)$ is said to be \emph{hierarchical} if
      for any $k \in \{1,\cdots, n\}$ $R_k|_{k-1}=R_{k-1}$.
  
\end{definition}



\begin{example}\label{ex:counter}
  Our running example is clearly a hierarchical model. Examples of non-hierarchical layered models can be achieved by removing some $0$-transitions depicted in Figure \ref{fig:intro2} (e.g. the one linking the named states $closed_0$ and $get\_access_0$). Observe that,  in this case,  one has                                                                              $@_{closed_0} \diamond_1 get\_access_0$ but $\neg @_{closed_0} \diamond_0 get\_access_0$.
\end{example}

\subsection*{Satisfaction} Let $M$ be a n-layered model. The satisfaction consists of a family of relations $\models^n=(\sat{k})_{k\in \{0,\cdots,n\}}$ defined, for each $w_r\in W^r$, $r\in \{0,\cdots, k\}$, $k\leq n$, such that $D_k(w_0,\cdots w_k)$, as follows: 
    \begin{itemize}
      \item $M_k,w_0,\cdots,w_k \sat{k} \varphi^b_{k-1}$ iff $M_{k-1},w_0,\cdots,w_{k-1} \sat{k-1} \varphi^b_{k-1}$
      \item $M_k,w_0,\cdots,w_k \sat{k} p_k$ iff $w_k \in V^\Prop_k(p_k,w_0,\cdots,w_{k-1})$
      \item $M_k,w_0,\cdots,w_k  \sat{k} i_k$ iff $w_k= V^\Nom_k(i_k)$ and  $D_k(w_0,\cdots,w_{k-1},V_k^\Nom(i_k))$
      \item $M_k,w_0,\cdots,w_k  \sat{k} \varphi_k \wedge \varphi'_k$ iff $M_k,w_0,\cdots,w_k \sat{k} \varphi_k$ and $M_k,w_0,\cdots,w_k  \sat{k}  \varphi'_k$ 
      \item  $M_k,w_0,\cdots,w_k  \sat{k} \neg \varphi_k$ iff it is false that $M_k,w_0,\cdots,w_k  \sat{k} \varphi_k$
            \item $M_k,w_0,\cdots,w_k  \sat{k} @_{i_k} \varphi_k$ iff $M_k,w_0,\cdots w_{k-1}, V^\Nom_k(i_k)  \sat{k} \varphi_k$  and $D_k(w_0,\cdots w_{k-1},V_k^\Nom(i_k))$
      \item $M_k,w_0,\cdots,w_k  \sat{k} \diamond_k \varphi_k$ iff $M,v_0,\cdots, v_k  \sat{k} \varphi_k$ for some $v_r\in W_r$, $r\in \{0,\cdots, k\}$,  such that \\  $(w_0,\dots, w_{k}) R_k (v_0,\cdots,v_{k})$.
    \end{itemize}


\begin{example} Let us illustrate this notion of satisfaction verifying the validity of $@_{idle_1} closed_{0}$ in  state $w_0^1$ of model $M$ in the running  example. For arbitrary $w_1,w_2$, 
\begin{eqnarray*}
  & & M_2,w_0,w_1,w_2  \sat{2} @_{idle_1} closed_{0}
  \just\Leftrightarrow{ defn. of $ \sat{2}$}
       M_1,w_0^1,w_1 \sat{1} @_{idle_1} closed_{0}
        \just\Leftrightarrow{ defn. of $ \sat{1}$}
        M_1,w_0^1,V_1^\Nom(idle_1) \sat{1} closed_{0}, \mbox{ and } D_2(w_0^1,V^\Nom_1(idle_1))=D_2(w_0^1,w_1^1)
        \just\Leftrightarrow{ defn. of $ \sat{1}$}
        V_0^\Nom(closed_0) = w_0^1 \mbox{ and } D_2(V_0^\Nom(closed_0),w_1^1)=D_2(w_0^1,w_1^1)
\end{eqnarray*}



\noindent
As a second illustration consider, 
\begin{eqnarray*}
  & & M_1,w_0^2,w_1^3  \sat{1} \diamond_1 get\_access_0 \wedge \neg \diamond_0 get\_access_0
  \just\Leftrightarrow{ defn. of $ \sat{1}$}
       M_1,w_0^2,w_1^3  \sat{1} \diamond_1\, get\_access_0 \mbox{ and }\mbox{ it is false that }  M_1,w_0^2,w_1^3  \sat{1} \diamond_0\, get\_access_0 
        \just\Leftrightarrow{ defn. of $ \sat{1}$}
        \mbox{there are }v_0,v_1 \mbox{ such that } (w_0^2,w_1^3)R_1(v_0,v_1) \mbox{ and } M_1,v_0,v_1  \sat{1} get\_access_0\\ & & 
        \mbox{ and }\\ & & 
        \mbox{ it is false that }
        \mbox{there is a }r_0 \mbox{ such that } (w_0^2)R_0(r_0) \mbox{ and } M_0,r_0  \sat{0} get\_access_0
        \just\Leftrightarrow{ defn. of $ \sat{1}$}
        \mbox{there are }v_0,v_1 \mbox{ such that } (w_0^2,w_1^3)R_1(v_0,v_1) \mbox{ and } v_0 = V_0^\Nom(warning_0) \\ & & \mbox{ (when $v_0=w_0^2$ and $v_1=w_1^4$)}
        \mbox{ and }\\ & & 
        \mbox{there is not a }r_0 \mbox{ such that } (w_0^2)R_0(r_0) \mbox{ and } r_0= V_0^\Nom(warning_0)
\end{eqnarray*}



\end{example}



\section{Basic modal theory}\label{sc:meh}
  
This section discusses three basic ingredients in a modal theory: the existence of a standard translation to first-order logic,  a notion of bisimulation and a modal invariance result. 


\subsection{Standard translation} 
  \label{sub:standard_translation}
  
Beyond the theoretical interest of this characterization, a standard translation  to first-order logic paves the way to the use of a number of tools to provide assistance and effective support for the refinement strategies suggested here.


    
   

\paragraph{Signature translation:} An $n$-layered signature $\Delta^n=(\Nom_n,\Prop_n)$ induces, for each $k\in \{0,\dots, n\}$, a first-order signature $(S^k,F^k,P^k)$ as follows:  
\begin{itemize}
  \item $S^k=\{S_0,\cdots,S_k\}$;
  \item $F^k$ is the $({S^k}^*,S^k)$-family of function symbols consisting of:
  \begin{itemize}
    \item for each $r\in \{0,\cdots, k\}$, $F^k_{\rightarrow S_r}=\{i_r| i_r\in \Nom_r\}$   
    \item and $F_{\omega \rightarrow S} = \emptyset$ for the other cases.
  \end{itemize} 
  \item $P^k$ is a ${S^k}^*$-family of predicate symbols such that for any $r\in \{0,\cdots,k\}$:
   \begin{itemize} 
    \item $P_{S_0,\cdots S_{r}}= \{D_r\}$
    \item $P_{S_1\cdots S_r}=\Prop_r$
    \item $P_{S_0,\cdots S_{r},S_0,\cdots S_{r}}=\{R_r\}$
    \item and $P_\omega = \emptyset$ for the other cases.
  \end{itemize} 
\end{itemize}



\paragraph{Models translation:}Let $M$ be a $\Delta^n$ model. For each $k\in \{0,\cdots,n\}$, the $(S^k,F^k,P^k)$-model $M^*_k$, corresponding to the translation of  $M_k$, is built as follows.
For each $r\in \{0,\cdots, k\}$:
\begin{itemize}
  \item ${M^*_k}_{S_r}=W_r$
  \item for each $i_r:\rightarrow S_r$,  ${M^*_k}_{i_r}= V_r^\Nom(i_r)$
  \item for any $p_r \in  P_{S_{r}}$, ${M^*_k}_{p_r}(w_0,\cdots w_r)=V^\Prop_r(p_r,w_0,\cdots,w_r)$
  \item  ${M^*_k}_{D_r}=D_r$
  \item  ${M^*_k}_{R_r}=R_r$
\end{itemize}

\paragraph{Sentences translation:}
The translation of sentences is recursively defined as follows:
~\\

  \begin{tabular}{rcll}
  $\ST^k_{x_0,\cdots,x_k}(\varphi^b_{k-1})$ & $=$ & $\ST^{k-1}_{x_0,\cdots,x_{k-1}}(\varphi^b_{k-1})$ & \\  
  $\ST^k_{x_0,\cdots,x_k}(p_k)$ & $=$ & $p_k(x_0,\cdots,x_k)$ & $p_k \in \Prop_k$ \\  
  $\ST^k_{x_0,\cdots,x_k}(i_k)$ & $=$ & $i_k=x_k$ & $i_k \in \Nom_k$ \\  
    $\ST^k_{x_0,\cdots,x_k}(@_{i_k}(\varphi_k))$ & $=$ & $D_k(x_0,\cdots,x_{k-1},i_k) \wedge \ST^k_{x_0,\cdots,x_{k-1},i_k}(\varphi_k)$ & $i_k \in \Nom_k$\\
  $\ST^k_{x_0,\cdots,x_k}(\diamond_k \varphi_k)$ & $=$ & $(\exists y_{0}, \cdots y_{k})\big(D_k(y_{0}, \cdots, y_k)\wedge$  & \\ & & $  R_k(x_0,\cdots,x_k,y_{0}, \dots, y_k)\wedge  \ST^k_{y_0,\dots y_k}(\varphi_k)\big)$ & \\  
    $\ST^k_{x_0,\cdots,x_k}(\varphi_k \wedge \varphi'_k)$ & $=$ & $\ST^k_{x_0,\cdots,x_k}(\varphi_k) \wedge \ST^k_{x_0,\cdots,x_k}(\varphi'_k)$ & \\  
  $\ST^k_{x_0,\cdots,x_k}(\neg \varphi_k)$ & $=$ & $\neg \ST^k_{x_0,\cdots,x_k}(\varphi_k)$ & \\  
  \end{tabular}
~\\

\begin{example}
\vspace{0.4cm}  
\begin{eqnarray*}
 \ST^1_{x_0,x_1}(@_{idle_1}closed_0)\; & =\; D_1(x_0,idle_1)\wedge \ST^1_{x_0,idle_1}(closed_0) \\
 & = D_1(x_0,idle_1)\wedge \ST^0_{x_0} (closed_0)\\
 & = D_1(x_0,idle_1)\wedge (closed_0 = x_0)
\end{eqnarray*}
\vspace{0.1cm}  
\end{example}



\begin{theorem}
 Let $M$ be a $n$-layered model of $\Delta^n$ and $\varphi_k$, $k\in\{0,\cdots,n\}$, a formula of $Fm_k(\Delta^n)$. Then,
\[M_k,w_0,\cdots,w_k  \sat{k} \varphi_k \; \mbox{ iff }\;  \overline{M_k}\models \ST^k_{x_0\cdots x_k} (\varphi_k)\]
where $\overline{M_k}$ is the $x_0,\cdots,x_k$-expansion of $M^*_k$ such that $\overline{M_k}_{x_r}=w_r$, for any $r\in \{0,\cdots, k\}$,
and $\models$ stands for the first order satisfaction relation.
 
\end{theorem}

\paragraph{Proof.} The proof is done by induction over the sentences and satisfaction structure. 
For $k=0$, the theorem boils down  to the corresponding result for usual standard translation for propositional hybrid logic  (see e.g. \cite{livro_brauner}). For the remaining cases:

\begin{description}

\item[Case of formulas $\varphi^b_{k-1}$]

\begin{eqnarray*}
  & & M_k,w_0,\cdots, w_k  \sat{k} \varphi^b_{k-1}
  \just\Leftrightarrow{ defn. of $ \sat{k}$}
       M_{k-1},w_0,\cdots, w_{k-1}  \sat{k-1} \varphi^b_{k-1}
        \just\Leftrightarrow{I.H.}
        \overline{M_{k-1}}\models \ST^{k-1}_{x_0,\cdots,x_{k-1}}(\varphi^b_{k-1})
        \just\Leftrightarrow{ defn. of $\ST^k$ and $\ST^{k-1}_{x_0,\cdots,x_{k-1}}(\varphi^b_{k-1})$ does not depend on $x_k$}
        \overline{M_k}\models \ST^k_{x_0,\cdots,x_k}(\varphi^b_{k-1})
\end{eqnarray*}

\item[Case of formulas $@_{i_k}\varphi$]


\begin{eqnarray*}
  & & M_k,w_0,\cdots, w_k  \sat{k} @_{i_k}\varphi
  \just\Leftrightarrow{ defn. of $ \sat{k}$}
        M_k,w_0,\dots, w_{k-1},V_k^\Nom(i_k)  \sat{k} \varphi \mbox{ and } D_k(w_0,\cdots,w_{k-1},V_k^\Nom(i_k))
        \just\Leftrightarrow{I.H. + defn. of $\overline{M_k}$}
        \overline{M_k}_{D_k}(\overline{M_k}_{x_0},\cdots,\overline{M_k}_{x_{k-1}},\overline{M_k}_{i_k}) \mbox{ and }\overline{M_k}\models \ST^k_{x_0,\cdots,x_{k-1}}(\varphi)
        \just\Leftrightarrow{$\models$ defn}
        \overline{M_k}\models D_k(x_0,\cdots,x_{k-1},i_k) \wedge \ST^k_{x_0,\cdots,x_{k-1},i_k}(\varphi)
        \just\Leftrightarrow{$\ST^k$ defn}
        \overline{M_k}\models \ST^k_{x_0,\cdots,x_{k}}(@_{i_k}\varphi)
\end{eqnarray*}

  \item[Case of formulas $i_k$]

\begin{eqnarray*}
  & & M_k,w_0,\cdots, w_k  \sat{k}i_k
  \just\Leftrightarrow{ defn. of $ \sat{k}$}
       w_k\in V_k^\Nom(i_k) \mbox{ and } D_k(w_0,\cdots,w_{k-1},V_k^\Nom(i_k))
        \just\Leftrightarrow{defn. of  $\models$}
        {M_k}_{i_k}=w_k \mbox{ and } M_{D_k}(w_0,\cdots,w_{k-1},{M_k}_{i_k})
        \just\Leftrightarrow{ defn. of $\overline{M_k}$}
        \overline{M_k}_{i_k}=\overline{M_k}_{x_k} \mbox{ and } \overline{M}_{D_k}(\overline{M}_{x_0},\cdots,\overline{M}_{x_{k-1}},\overline{M_k}_{i_k})
        \just\Leftrightarrow{ defn. of $\models$}
        \overline{M_k}\models {i_k}=x_k \wedge D_k(x_0,\cdots, x_{k-1},i_k)
        \just\Leftrightarrow{ defn. of $\ST^k$}
        \overline{M_k}\models \ST^k_{x_0,\cdots,x_k}(i_k)
\end{eqnarray*}

  \item[Case of formulas $p_k$]

\begin{eqnarray*}
  & & M_k,w_0,\cdots, w_k  \sat{k}p_k
  \just\Leftrightarrow{ defn. of $ \sat{k}$}
       w_k\in V_k^\Prop(p_k,w_0,\cdots,w_{k-1}) 
        \just\Leftrightarrow{defn. of  $M_k$}
        {M_k}_{p_k}(w_0,\cdots,w_k)
        \just\Leftrightarrow{ defn. of $\overline{M_k}$ + defn. $\models$}
        \overline{M_k}\models p_k(x_{0},\cdots,x_k)
        \just\Leftrightarrow{ defn. of $\ST^k$}
        \overline{M_k}\models \ST^k_{x_0,\cdots,x_k}(p_k)
\end{eqnarray*}

  \item[Case of formulas $\diamond_k\varphi$]

\begin{eqnarray*}
  & & M_k,w_0,\cdots, w_k  \sat{k} \diamond_k\varphi
  \just\Leftrightarrow{ defn. of $ \sat{k}$}
        M_k,v_0,\dots,v_k  \sat{k} \varphi  \mbox{ for some } v_r \in W_r, \mbox{ such that }
        (w_0,\dots, w_{k})R_k (v_0,\dots,v_k)
        \just\Leftrightarrow{defn. of  $\overline{M_k}$ + FOL semantics + $R_k\subseteq D_k \times D_k$}
        \overline{M_k}\models (\exists y_0,\cdots y_k)\big(D(x_0,\cdots,x_{k-1},y_k,\cdots,y_n)\wedge R_k(x_0,\cdots, x_k,y_0 \cdots,y_{k}) \wedge \ST^k_{y_0,\cdots, y_k}(\varphi)\big)
        \just\Leftrightarrow{ defn. of $\ST$}
       \overline{M_k}\models\ST^k_{x_0,\cdots, x_k}(\diamond_k \varphi)
\end{eqnarray*}

\end{description}
The proof for the boolean connectives is straightforward. \qed

\subsection{Bisimulation and modal invariance} 
\label{sub:bisimulation}

Bisimulation is the main conceptual tool to compare transition systems. The originality in the definition below is the way the layered structured is taken into account in the \emph{zig-zag} conditions. This is illustrated  in Figure \ref{fig:bisimulation} below. The remaining components are, as expected, completely standard in hybrid logic. Note, for example, the condition imposed on nominals which makes bisimilarity a quite fine-grained equivalence. Back to Figure \ref{fig:bisimulation} this condition  forces us not to consider  nominals in the transition structures represented there.

\begin{definition}[$n$-layered bisimulation] \label{def:nbis}
    Let $M$ and $M'$ be two $n$-layered models over the signature $\Delta^n=(\Prop_k,\Nom_k)_{k\in \{0,\cdots,n\}}$. A family of relations \[B=(B_k\subseteq D_k \times D'_k)_{k\in \{0,\cdots,n\}}\]
is a $n$-layered bisimulation ($n$-bisimulation for short) if, for any $k\in\{0,\cdots, n\}$,
 whenever \[(w_0,\cdots, w_k)B_k({w'}_0,\cdots, {w'}_k)\] 
we have that:
\begin{description}
    \item[($ATOM_k$)]
        \begin{itemize}
        \item[1.] for each $p_k\in \Prop_k$, 
        \begin{itemize}
          \item[i.] if $k=0$ we have $w_0\in V^\Prop_0(p_0) \mbox{ iff } {w'}_0\in {V'}^\Nom_0(p_0)$;
          \item[ii.] otherwise, $w_k\in V_k(p_k,w_0,\cdots,w_{k-1}) \mbox{ iff } {w'}_k\in V'_k(p_k,w'_0,\cdots, {w'}_{k-1})$.
        \end{itemize} 
      \item[2.] for each $i_k\in \Nom_k$, 
      \begin{itemize}
         \item[i.] if k=0, $(V_0^\Nom(i_0))B_k({V'}_0^\Nom(i_0))$; \\otherwise, $(w_0,\cdots, w_{k-1},V_k^\Nom(i_k))B_k({w'}_0,\cdots, {w'}_{k-1},{V'}_k^\Nom(i_k))$, 
         \item[ii.] $w_k= V_k(i_k) \mbox{ and } D_k(w_0,\cdots, w_{k-1},{V}_{k}^{\Nom}(i_k))  \mbox{ iff } {w'}_k=V'_k(i_k)\mbox{ and }$ \\ $ D'_k(w'_0,\cdots, w'_{k-1},{V'}_{k}^ {\Nom}(i_k)) $
       \end{itemize} 
    \end{itemize}
  
\item[($ZIG_k$)] 
for any $v_r\in W_r$, $r\in \{0,\cdots, k\}$,  such that $(w_{0},\cdots,w_k)R_k (v_{0},\cdots,v_k)$, there are $v'_r\in {W'}_r$, $r\in \{0,\cdots, k\}$, such that 
$({w'}_{0},\cdots,{w'}_k){R'}_k ({v'}_{0},\cdots{v'}_k)$ and 
\begin{equation}\label{eq:bis}
(v_{0},\cdots,v_k)B_k({v'}_{0},\cdots, v'_k)  
\end{equation}

\item[($ZAG_k$)] for any ${v'}_r\in {W'}_r$, $r\in \{0,\cdots, k\}$,  such that
$({w'}_{0},\cdots,{w'}_k){R'}_k ({v'}_{0},\cdots,{v'}_k)$, there are ${v}_r\in {W}_r$, $r\in \{0,\cdots, k\}$, such that
$({w}_{0},\cdots,{w}_k){R}_k ({v}_{0},\cdots,{v}_k)$ satisfy (\ref{eq:bis})
\end{description}
 \end{definition}

\begin{lemma}
In the conditions of the previous definition, $B[k]\subseteq D_k \times D_k$ is a $k$-bisimulation between $M_k$ and $M'_k$,  
for any $k\in\{0,\cdots,n\}$.
\end{lemma}

\begin{theorem}\label{th:bisimk}
Let $M$ and $M'$ be two $n$-layered models over the signature $\Delta^n$ and
$B$ a $n$-layered bisimulation. Then, for $k\in\{0,\cdots,n\}$ and  $w_k \in W_k$, ${w'}_k \in {W'}_k$,  such that $(w_{0},\cdots, w_k)B_k(w'_{0},\cdots, w'_k)$
and any $\varphi_k\in SFm(k,\Delta^n)$,
\[M_k,w_0,\cdots, w_k  \sat{k} \varphi_k \; \mbox{ iff }\;  M'_k,w'_0,\cdots, w'_k \sat{k} \varphi_k\]
\end{theorem}

\paragraph{Proof.} The proof is done by induction over the sentences and satisfaction structure. 
For $k=0$, the theorem boils down  to the corresponding modal invariance result for propositional hybrid logic (see e.g. \cite{livro_brauner}). For the remaining cases:

\begin{description}
  \item[Case of formulas $\diamond_k\varphi_k$]
\begin{eqnarray*}
  & & M_k,w_0,\cdots, w_k  \sat{k} \diamond_k\varphi_k
  \just\Leftrightarrow{ defn. of $ \sat{k}$}
        M_k,v_0,\dots,v_k  \sat{k} \varphi_k \mbox{ for some } v_r \in W_r, r\in \{0,\cdots,k\}
        \\ & & \mbox{ such that }
        (w_0,\cdots, w_{n})R_k (v_0,\cdots, v_k)
        \just\Leftrightarrow{Step ($\star$)}
        M'_k,v'_0,\cdots, v'_k  \sat{k} \varphi_k \mbox{ for some }v'_r \in {W'}_r, r\in \{0,\cdots,k\}
        \\ & & 
        \mbox{ such that }
        (w'_0,\cdots, w'_{n})R'_k (v'_0,\cdots, v'_k)
        \just\Leftrightarrow{ defn. of $ \sat{k}$}
       M'_k,w'_0,\cdots, w'_k  \sat{k} \diamond_k\varphi_k
\end{eqnarray*}
\noindent \textbf{Step ($\star$):} For the up-down implication, the existence of $v'_0,\cdots, v'_k$ such that \\ $(w_0,\cdots, w_{k})B_k (v_0,\cdots,v_{k})$ is assured by  the recursive application of ($ZIG_k$). Then conclude by I.H. The down-up implication is  proved similarly, but resorting to the  ($ZAG_k)$ condition.


  \item[Case of formulas $@_{i_k}\varphi_k$]


\begin{eqnarray*}
  & & M_k,w_0,\cdots, w_k  \sat{k} @_{i_k}\varphi_k
  \just\Leftrightarrow{ defn. of $ \sat{k}$}
        M_k,w_0,\cdots, w_{k-1},V_k^\Nom(i_k)  \sat{k} \varphi_k \mbox{ and } D_k(w_0,\cdots,w_{k-1},V_k^\Nom(i_k))
        \just\Leftrightarrow{ $(w_0,\cdots,w_k)B_k(w'_0,\cdots,w'_k)$ + $ATOM_k$ (2.i.) + I.H.}
        M'_k,w'_0,\cdots, w'_{k-1},{V'}_k^\Nom(i_k)  \sat{k} \varphi_k \mbox{ and } D'_k(w'_0,\cdots,w'_{k-1},{V'}_k^\Nom(i_k))
        \just\Leftrightarrow{ defn. of $ \sat{k}$}
       M'_k,w'_0,\cdots, w'_k  \sat{k} @_{i_k}\varphi_k  
\end{eqnarray*}

  \item[Case of formulas $i_k$]

\begin{eqnarray*}
  & & M_k,w_0,\cdots, w_k  \sat{k} i_k 
  \just\Leftrightarrow{ defn. of $ \sat{k}$}
        w_k = V^\Nom_k(i_k) \mbox{ and } D_k(w_0,\cdots,w_{k-1},V_k^\Nom(i_k))
        \just\Leftrightarrow{$ATOM_k$ (2.i.)}
        w'_k = V'^\Nom_k(i_k) \mbox{ and } D'_k(w'_0,\cdots,w'_{k-1},{V'}_k^\Nom(i_k))
        \just\Leftrightarrow{ defn. of $ \sat{k}$}
       M'_k,w'_0,\cdots, w'_k  \sat{k} i_k 
\end{eqnarray*}

\end{description}
The proof for propositions is analogous and for the boolean connectives is straightforward. 

\qed

\medskip\noindent
Definition \ref{def:nbis} boils down to the following version of bisimulation for hierarchical systems:

\begin{definition}[$n$-hierarchical bisimulation]
    Let $M$ and $M'$ be two $n$-hierarchical models over $\Delta^n=(\Prop_k,\Nom_k)_{k\in \{0,\cdots,n\}}$. A $n$-\emph{hierarchical bisimulation} consists of a relation $B\subseteq D_n \times D'_n$
such that, for any $k\in \{0,\cdots, n\}$ and for any $w_r \in W_r$ and $w'_r \in {W'}_r$, $r\in\{0,\dots,k\}$ such that
 $(w_0,\cdots, w_n)B({w'}_0,\cdots, {w'}_n)$, we have that:
\begin{description}
\item[($ATOM$)]
    \begin{itemize}
        \item[1.] for each $p_k\in \Prop_k$,
        \begin{itemize}
           \item[i.]if $k=0$ we have $w_0\in V^\Prop_0(p_0) \mbox{ iff } {w'}_0\in {V'}^\Nom_0(p_0)$;
           \item[ii.] otherwise, $w_k\in V_k(p_k,w_0,\cdots,w_{k-1}) \mbox{ iff } {w'}_k\in V'_k(p_k,w'_0,\cdots, {w'}_{k-1})$
         \end{itemize}  
      \item[2.] for each $i_k\in \Nom_k$, 
      \begin{itemize}
         \item[i.] if k=0, $(V_0^\Nom(i_0))B|_0({V'}_0^\Nom(i_0))$; \\ otherwise, $(w_0,\cdots, w_{k-1},V_k^\Nom(i_k))B|_k({w'}_0,\cdots, {w'}_{k-1},{V'}_k^\Nom(i_k))$, 
         \item[ii.] $w_k= V_k(i_k) \mbox{ and } D_k(w_0,\cdots, w_{k-1},{V}_{k}^{\Nom}(i_k))  \mbox{ iff } {w'}_k=V'_k(i_k)\mbox{ and } \\ D'_k(w'_0,\cdots, w'_{k-1},{V'}_{k}^ {\Nom}(i_k)) $
       \end{itemize} 
    \end{itemize}

  
\item[($ZIG$)] 
for any $v_r\in W_r$, $r\in \{0,\cdots, n\}$,  such that $(w_{0},\cdots,w_n)R (v_{0},\cdots,v_n)$, there are $v'_r\in {W'}_r$, $r\in \{0,\cdots, n\}$, such that
$({w'}_{0},\cdots,{w'}_n){R'} ({v'}_{0},\cdots{v'}_n)$ and 
\begin{equation}\label{eq:bis2}
(v_{0},\cdots,v_n)B({v'}_{0},\cdots, v'_n)  
\end{equation}

\item[($ZAG$)] for any ${v'}_r\in {W'}_r$, $r\in \{0,\dots, n\}$,  such that
$({w'}_{0},\cdots,{w'}_n){R'} ({v'}_{0},\cdots,{v'}_n)$, there are ${v}_r\in {W}_r$, $r\in \{0,\cdots, n\}$, such that
$({w}_{0},\cdots,{w}_n){R} ({v}_{0},\cdots,{v}_n)$ satisfying (\ref{eq:bis2})
\end{description}
 \end{definition}

\begin{figure}[ht!]
\begin{center}
   \includegraphics[width=0.5\textwidth,natwidth=610,natheight=642]{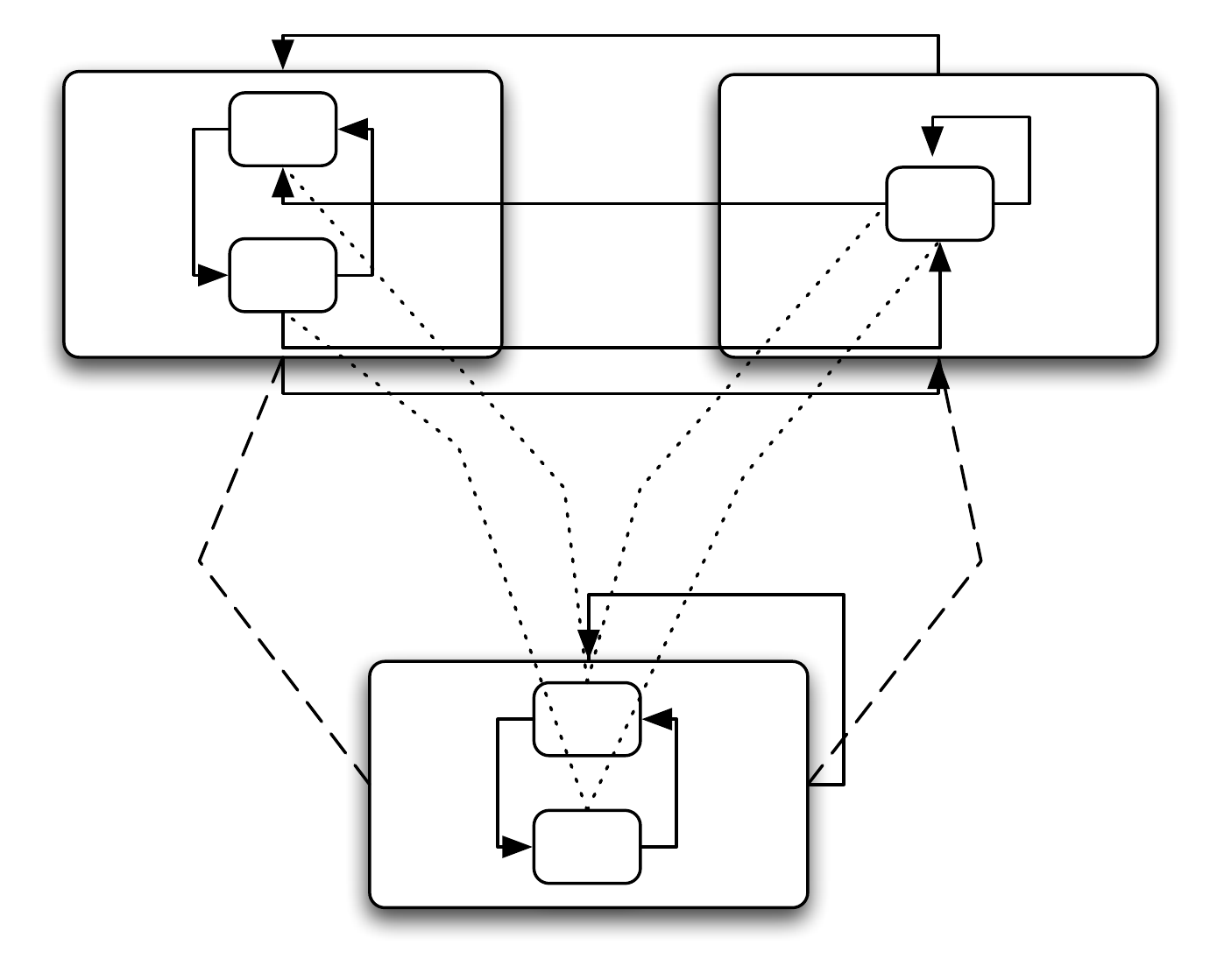}
\end{center}
\caption{A 2-hierarchical bisimulation.}
\label{fig:bisimulation}
\end{figure}

\begin{corollary}\label{cor:bisimk}
Let $M$ and $M'$ be two $n$-hierarchical models over the signature $\Delta^n$ and
$B$ an $n$-hierarchical bisimulation. Then for any $w_k \in W_k$ and $w'_k \in {W'}_k$, $k\in\{0,\cdots,n\}$ such that $(w_{0},\cdots, w_n)B({w'}_{0},\cdots, {w'}_n)$
and for any $\varphi\in Fm_n(\Delta^n)$,
\[M,w_0,\cdots, w_n  \sat{n} \varphi\; \mbox{ iff }\; M',{w'}_0,\cdots, {w'}_n  \sat{n} \varphi\]
\end{corollary}

\noindent \textbf{Proof.} First of all observe that a $n$-hierarchical bisimulation $B\subseteq D_n \times {D'}_n$ induces a $n$-layered bisimulation $\overline{B}$ by taking, for each $k\in \{0,\cdots,n\}$, the component $\overline{B}_k=B|_k$. 

Since $\varphi\in Fm(\Delta^n)$ then $\varphi\in Fm_n(\Delta^n)$. Hence since $M,w_0,\dots,w_n  \sat{n} \varphi$ iff $M_k,w_0,\cdots,w_k  \sat{k} \varphi$. Moreover, since $M$ and $M'$ are hierarchical, we have that $(w_0,\cdots,w_k)B[k](w'_0,\cdots,w'_k)$.  By Theorem~\ref{th:bisimk} to achieve at $M'_k,w'_0,\cdots,w'_k  \sat{k} \varphi$. Hence $M',w'_0,\cdots,w'_n  \sat{n} \varphi$.
\qed

\section{Refinement}\label{sc:refine}
\subsection{Simulation} 
\label{sub:simulation}

As usual, simulation entails a notion of refinement. However, as done with bisimulations in section \ref{sc:meh}, a distinction is made between the general notion of $n$-layered simulation and more well-behaved' \emph{$n$-hierarchical} one. Definitions and results are as expected. Thus,


\begin{definition}[$n$-layered simulation]
    Let $M$ and $M'$ be two $n$-layered models over the signature $\Delta^n=(\Prop_k,\Nom_k)_{k\in \{0,\cdots,n\}}$. A family of relations \[S=(S_k\subseteq S_k \times S'_k)_{k\in \{0,\cdots,n\}}\]
is a $n$-layered simulation from $M$ to $M'$ if for any $k\in\{0,\cdots, n\}$
and
for any $w_r \in W_r$ and $w'_r \in {W'}_r$, $r\in\{0,\cdots,k\}$, 
    whenever $(w_0,\cdots, w_k)S_k({w'}_0,\cdots, {w'}_k)$, 
we have that:
\begin{description}
\item[($ATOM_k$)]\,
    \begin{itemize}
        \item[1.] for each $p_k\in \Prop_k$,
        \begin{itemize}
           \item[i.]when $k=0$, if $w_0\in V^\Prop_0(p_0) \mbox{ then } {w'}_0\in {V'}^\Nom_0(p_0)$;
           \item[ii.] otherwise, $w_k\in V_k(p_k,w_0,\cdots,w_{k-1}) \mbox{ implies that } {w'}_k\in V'_k(p_k,w'_0,\cdots, {w'}_{k-1})$
         \end{itemize}  
      \item[2.] for each $i_k\in \Nom_k$, 
      \begin{itemize}
         \item[i.] if k=0, $(V_0^\Nom(i_0))S_k({V'}_0^\Nom(i_0))$; \\ otherwise, $(w_0,\cdots, w_{k-1},V_k^\Nom(i_k))S_k({w'}_0,\cdots, {w'}_{k-1},{V'}_k^\Nom(i_k))$, 
         \item[ii.] $w_k= V_k(i_k) \mbox{ and } D_k(w_0,\cdots, w_{k-1},{V}_{k}^{\Nom}(i_k))  \mbox{ implies that } {w'}_k=V'_k(i_k) \\ \mbox{ and } D'_k(w'_0,\cdots, w'_{k-1},{V'}_{k}^ {\Nom}(i_k)) $
       \end{itemize} 
    \end{itemize}

\item[($ZIG_k$)] 
for any $v_r\in W_r$, $r\in \{0,\cdots, k\}$,  such that $(w_{0},\cdots,w_k)S_k (v_{0},\cdots,v_k)$, there are $v'_r\in {W'}_r$, $r\in \{0,\cdots, k\}$, such that
$({w'}_{0},\cdots,{w'}_k){R'}_k ({v'}_{0},\cdots{v'}_k)$ and 
$(v_{0},\cdots,v_k)S_k({v'}_{0},\cdots, v'_k) $ 
\end{description}
 \end{definition}

\begin{theorem}\label{th:refk}
Let $M$ and $M'$ be  $n$-layered models over the signature $\Delta^n$ and
$S$ an $n$-layered simulation from $M$ to $M'$. Then for any $w_k \in W_k$ and ${w'}_k \in {W'}_k$, $k\in\{0,\cdots,n\}$ such that \\ $(w_{0},\cdots, w_k)S_k(w'_{0},\cdots, w'_k)$
and for any $\varphi_k\in SFm^+(k,\Delta^n)$,
\[M_k,w_0,\cdots, w_k  \sat{k} \varphi_k\; \mbox{ implies that } \; M'_k,w'_0,\cdots, w'_k  \sat{k} \varphi_k\]
\end{theorem}
\noindent \textbf{Proof.} Straightforward from Theorem \ref{th:bisimk}.\qed
\medskip

\noindent
The definition  specialises to one for $n$-hierarchical systems.

\begin{definition}[$n$-hierarchical simulation]
    Let $M$, $M'$ be two $n$-hierarchical models over \\ $\Delta^n=(\Prop_n,\Nom_n)$. A $n$-\emph{hierarchical simulation} consists of a relation $S\subseteq D_n \times D'_n$
 such that, for any $k\in \{0,\cdots, n\}$ and for any $w_r \in W_r$ and $w'_r \in {W'}^r$, $r\in\{0,\dots,k\}$,
whenever $(w_0,\cdots, w_n)S({w'}_0,\cdots, {w'}_n)$,
we have that
\begin{description}
\item[($ATOM$)]\,
    \begin{itemize}
        \item[1.] for each $p_k\in \Prop_k$,
        \begin{itemize}
           \item[i.]when $k=0$, if $w_0\in V^\Prop_0(p_0) \mbox{ then } {w'}_0\in {V'}^\Nom_0(p_0)$;
           \item[ii.] otherwise, $w_k\in V_k(p_k,w_0,\cdots,w_{k-1}) \mbox{ implies that } {w'}_k\in V'_k(p_k,w'_0,\cdots, {w'}_{k-1})$
         \end{itemize}  
      \item[2.] for each $i_k\in \Nom_k$, 
      \begin{itemize}
         \item[i.] if k=0, $(V_0^\Nom(i_0))S|_0({V'}_0^\Nom(i_0))$; \\ otherwise, $(w_0,\cdots, w_{k-1},V_k^\Nom(i_k))S|_k({w'}_0,\cdots, {w'}_{k-1},{V'}_k^\Nom(i_k))$, 
         \item[ii.] $w_k= V_k(i_k) \mbox{ and } D_k(w_0,\cdots, w_{k-1},{V}_{k}^{\Nom}(i_k))  \mbox{ implies that }$\\$ {w'}_k=V'_k(i_k)\mbox{ and } D'_k(w'_0,\cdots, w'_{k-1},{V'}_{k}^ {\Nom}(i_k)) $
       \end{itemize} 
    \end{itemize}


\item[($ZIG$)] 
for any $v_r\in W_r$, $r\in \{0,\cdots, n\}$,  such that $(w_{0},\dots,w_n)S (v_{0},\cdots,v_n)$, there are $v'_r\in {W'}_r$, $r\in \{0,\cdots, n\}$, such that
$({w'}_{0},\cdots,{w'}_n){R'} ({v'}_{0},\cdots{v'}_n)$ and 
$(v_{0},\dots,v_n)S({v'}_{0},\dots, v'_n)  $

\end{description}
 \end{definition}

%
%

Given two $n$-layered models $M$ and $M'$, we say that $M'$ $l$-simulates $M$, in symbols $M\HR_l M'$ if there exists a  total horizontal $n$-layered simulation $S$ from $M$ to $M'$.
Analogously, given two $n$-hierarchical models $M$ and $M'$, we say that $M'$ $h$-simulates $M$, in symbols $M\HR_{h} M'$, if there exists a  total $n$-hierarchical simulation $S$ from $M$ to $M'$.

\begin{example}\label{ex:href}
Back to our running example, suppose now that, to meet an additional  safety requirement, it is imposed that, whenever blocked, the strongbox has  to be reset under some specific administrative permissions. 
\begin{figure}[ht!]
\begin{center}
   \includegraphics[width=0.9\textwidth,natwidth=610,natheight=642]{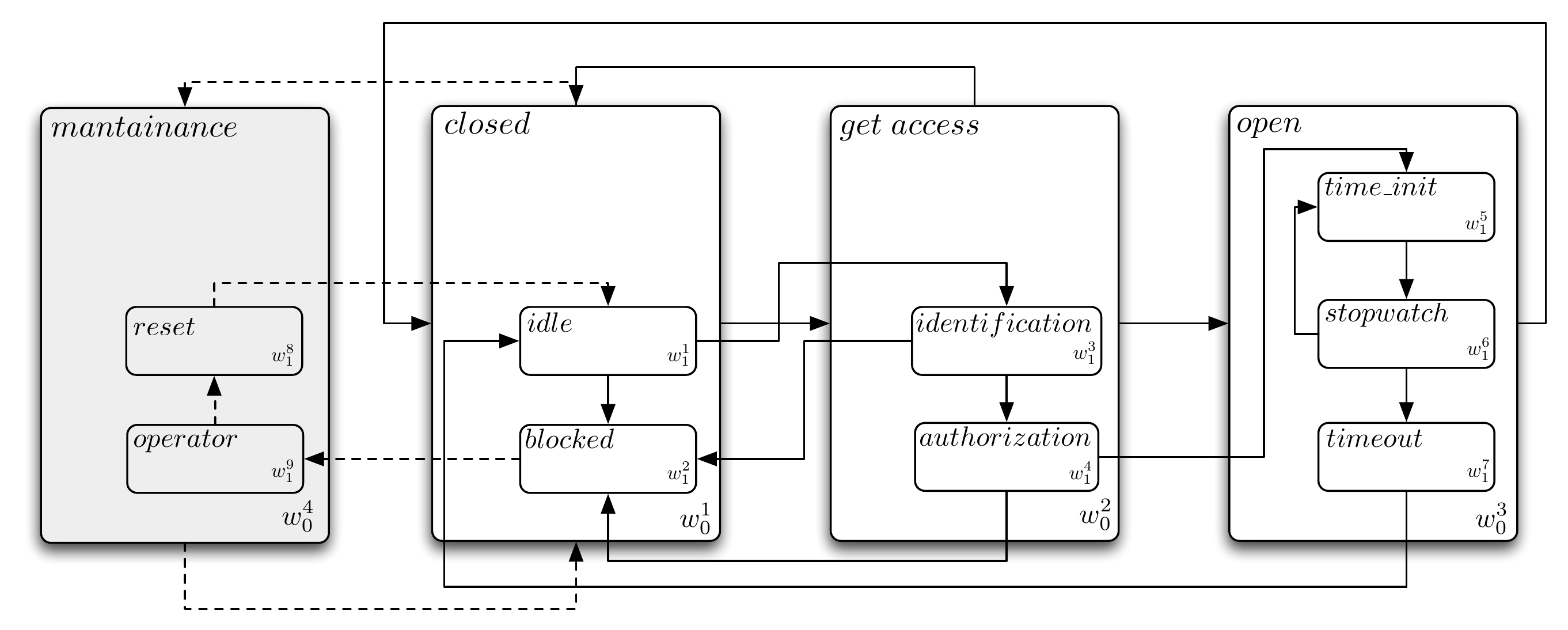}
\end{center}
\caption{Strongbox with administrative reset.}
\label{fig:href}
\end{figure}
\noindent This scenario is depicted in Figure~\ref{fig:href}. Clearly, this updated model  hierarchically simulates the original one in Figure~\ref{fig:intro2}. Actually the former model is a sub-model of the latter.

\end{example}

\begin{corollary}\label{cor:refk}
Let $M$ and $M'$ be two $n$-hierarchical models over the signature $\Delta^n$ and
$S$ an $n$-hierarchical simulation. Then for $k\in\{0,\cdots,n\}$, $w_k \in W_k$ and $w'_k \in {W'}_k$,  such that \\$(w_{0},\cdots, w_n)S({w'}_{0},\cdots, {w'}_n)$
and for any $\varphi\in Fm^+_n(\Delta^n)$,
\[M,w_0,\cdots, w_n  \sat{n} \varphi \mbox{ implies that } M',{w'}_0,\cdots, {w'}_n \sat{n} \varphi\]
\end{corollary}

\noindent \textbf{Proof.} Straightforward from Theorem~\ref{th:refk}. \qed

\subsection{Refinement}\label{sc:vf}

Finally, we have all ingredients to define refinement, distinguishing again the general $n$-layered case from the $n$-hierarchical one.
  

\begin{definition}[Layered refinement]
Let $\Delta^{n+k}$ be a $n+k$-layered signature and $\Delta^n=\Delta^{n+k}[n]$.
Let $M=(W^n,D^n,R^n,V^n)\in\Mod^n (\Delta^n)$ and $N=({W'}^{n+k},{D'}^{n+k},{R'}^{n+k},{V'}^{n+k})\in\Mod^{n+k} (\Delta^{n+k})$. We say that $N$ is an \emph{layered refinement} of $M$, in symbols
    $\refinesl{M}{N}$, whenever $M\HR_l N_n$.
  
\end{definition}

\begin{definition}[Hierarchical refinement]
Let $\Delta^{n+k}$ be a $n+k$-hierarchical signature and $\Delta^n=\Delta^{n+k}[n]$.
Let $M=(W^n,D^n,R^n,V^n)\in\Mod^n (\Delta^n)$ and $N=({W'}^{n+k},{D'}^{n+k},{R'}^{n+k},{V'}^{n+k})\in\Mod^{n+k} (\Delta^{n+k})$. We say that $N$ is an \emph{hierarchical refinement} of $M$, in symbols
    $\refinesh{M}{N}$, whenever $M\HR_h N_n$.
  
\end{definition}

\begin{theorem}
Let $M=(W^n,D^n,R^n,V^n)$ and $N=({W'}^{n+k},{D'}^{n+k},{R'}^{n+k},{V'}^{n+k})$ two layered models of $\Delta^n$ and $\Delta^{n+k}$ respectively. Then, if $\refinesl{M}{N}$, we have for any $\varphi \in SFm^+(k,\Delta^n)$ and for any $w_r\in {W'}_{r}$, $r\in \{n+1,\cdots, n+k\}$, 
\[M,w_0,\cdots,w_n \sat{n} \varphi \mbox{ implies that } N,w_0,\cdots,w_n,\cdots, w_{n+k} \sat{n+k} \varphi\]
\end{theorem}
\noindent \textbf{Proof.} Straightforward  from  Corollary~\ref{cor:refk}. \qed

\begin{theorem}

Let $M=(W^n,D^n,R^n,V^n)$ and $N=({W'}^{n+k},{D'}^{n+k},{R'}^{n+k},{V'}^{n+k})$ two hierarchical models of $\Delta^n$ and $\Delta^{n+k}$ respectively. Then, if $\refinesh{M}{N}$, we have for any $\varphi \in Fm^+(\Delta^n)$ and for any $w_r\in {W'}_{r}$, $r\in \{n+1,\cdots, n+k\}$, 
\[M,w_0,\cdots,w_n \sat{n} \varphi \mbox{ implies that } N,w_0,\cdots,w_n,\cdots, w_{n+k}\ \sat{n+k} \varphi\]
  
\end{theorem}
\noindent \textbf{Proof.} Straightforward  from  Theorem~\ref{cor:refk}. \qed

\begin{example}

It is easy to show that the model considered in Example~\ref{ex:href} is a $2$-hierarchical refinement of the one presented in Figure~\ref{fig:intro2}. Actually, its $1$-level restriction simulates the model presented in Figure~\ref{fig:intro1}, as illustrated in  Figure~\ref{fig:href0}.
\begin{figure}[ht!]
\begin{center}
 \includegraphics[width=0.8\textwidth,natwidth=610,natheight=642]{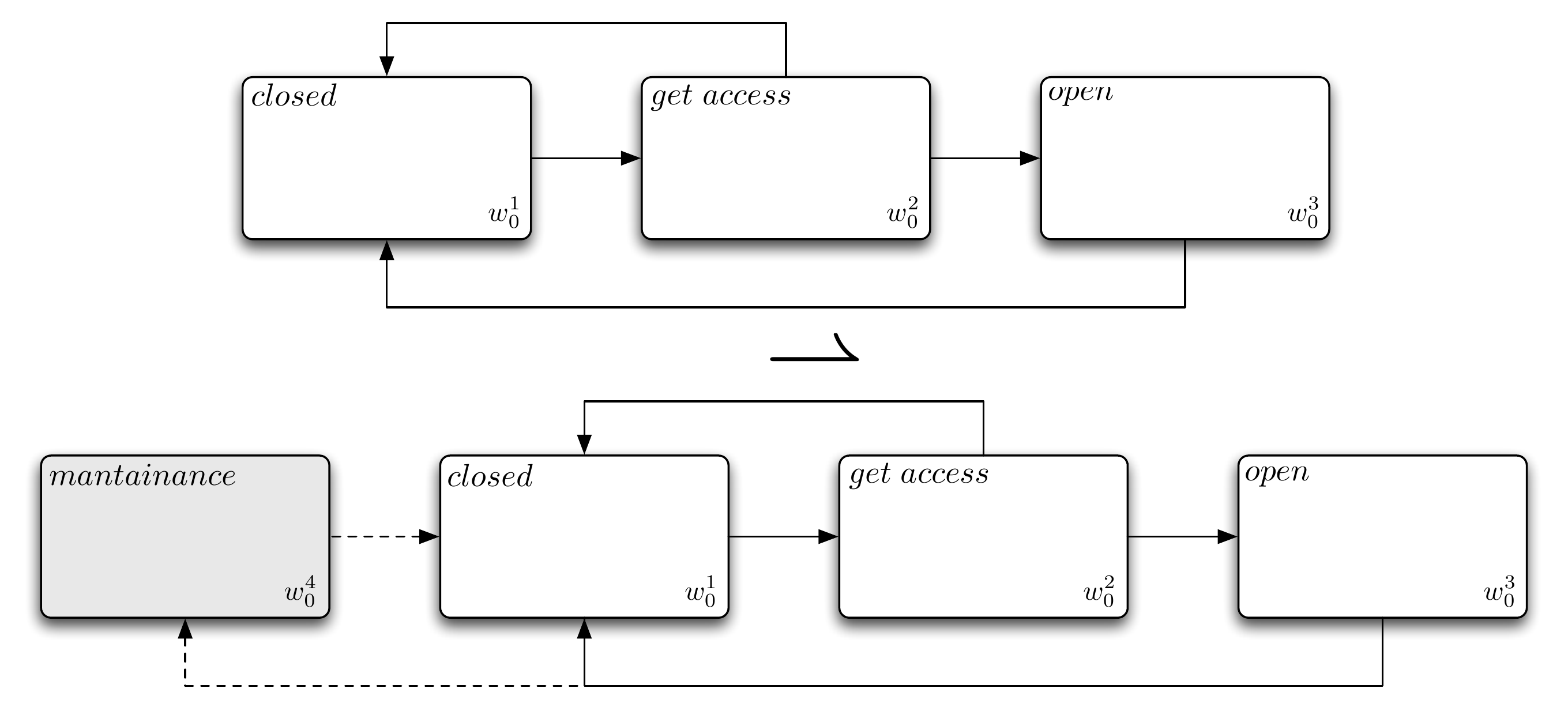}
\end{center}
\caption{An hierarchical refinement.}
\label{fig:href0}
\end{figure}
\end{example}

\begin{figure}[ht!]
\begin{center}
  \includegraphics[width=1\textwidth,natwidth=610,natheight=642]{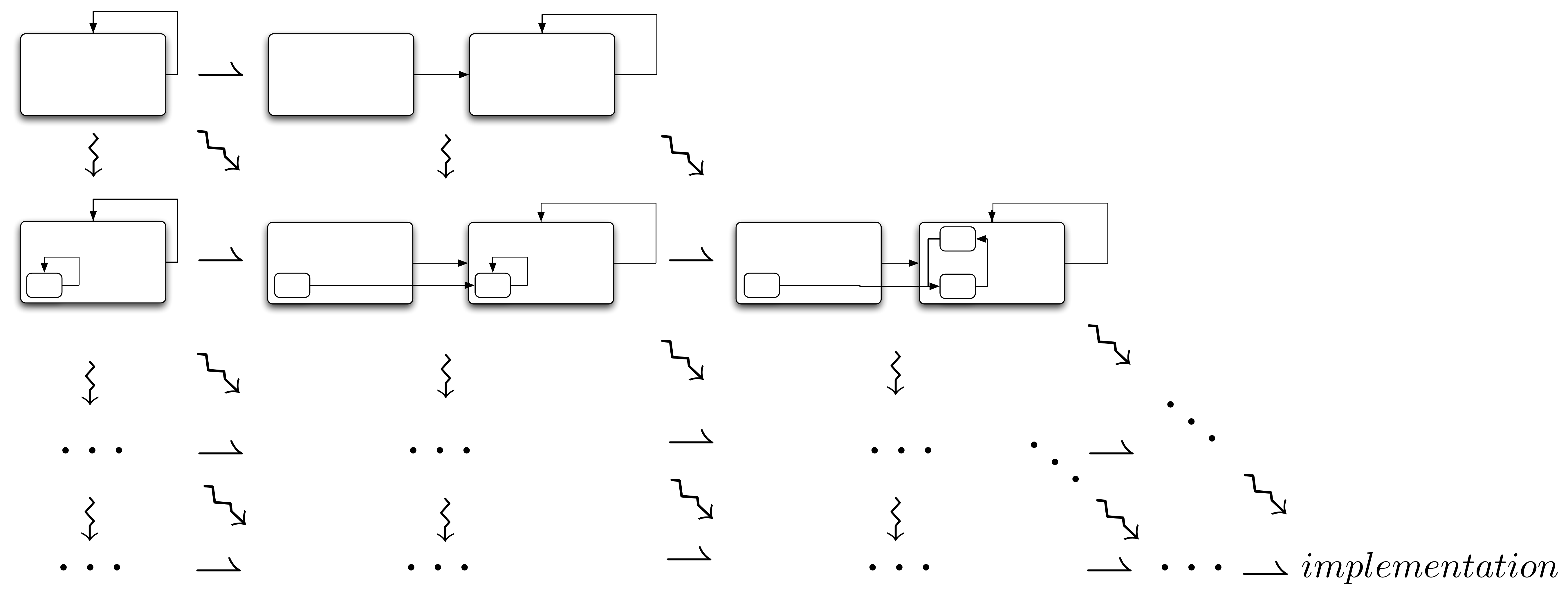}
\end{center}
\caption{A (hierarchical) stepwise refinement process.}
\label{fig:refinement}
\end{figure}

\noindent
Finally, Figure \ref{fig:refinement} illustrates the associated stepwise refinement process in which simulation steps combine with refinements until reaching a suitable implementation. Note that the strictly vertical arrows correspond to hierarchical steps along which, up to given level, the original and the refined transition systems are bisimilar. The diagonal arrows represent proper simulations between them.

\section{Conclusions and further work} 
\label{sc:tim}

The paper introduced  a hybrid modal logic for reasoning about k-layered transition systems and support horizontal and hierarchical refinement. The logic is expressive enough to capture  different forms of intra- and inter-level transitions present in most formalisms used in software specification and analysis with an  hierarchical flavour, spanning from D. Harel's \emph{statecharts} \cite{DBLP:journals/scp/Harel87} to the mobile \emph{ambients} \cite{CardelliG98} of A. Gordon and L. Cardelli,
Actually this work is rooted on the authors' previous study of what was called \emph{hybrid hierarchical logic}, $\HHL$, in reference \cite{aiml14} and, although being much more restrictive in the sort of expressible transitions, represented a first step in characterising a logic for hierarchical structures. Indeed  $\HHL$ arises from building a extra hybrid level (with new sets of nominals and modalities) on top   of standard, propositional hybrid logic. This  process, in full generality,  is called  \emph{hybridisation} \cite{madeirathesis,calco,paperdiaco} and consists of taking an arbitrary logic, framed as an institution \cite{ins} and systematically developing on top of it the syntax and semantic features of hybrid logic. Refinement in hybridised logics was studied by the authors in \cite{DBLP:journals/fac/MadeiraMBH15}.

The development of suitable notions of both horizontal and hierarchical refinements is one of the paper's contributions.  Current work  is therefore mainly concerned with proof-of-concept applications, namely the study  of variants of $k$-layered logics devoted to specific approaches in software engineering design. For instance, in a recent  institutional rendering of UML \cite{DBLP:conf/birthday/KnappMR15}, the formalisation of UML state-machines  leaves out  hierarchical states (see \cite{DBLP:journals/corr/KnappMR14}), a limitation that may be addressed in our framework. Other future research directions are concerned with decidability,  the development of a calculus and proof support.

\noindent
{\small
 {\bf Acknowledgements}.  
This work is funded by ERDF - European Regional Development Fund, through the COMPETE Programme, and by National Funds through FCT within project PTDC/EEI-CTP/4836/2014. A. Madeira is supported by the FCT grant SFRH/BPD/103004/2014. Finally, M. Martins is supported by FCT project \\ UID/MAT/04106/2013 at CIDMA and the EU FP7 Marie Curie PIRSES-GA-2012-318986 project GeT-Fun: Generalizing Truth-Functionality.}









\bibliographystyle{eptcs}

\end{document}